\documentclass[aps,prl,twocolumn,showpacs,%
	amsmath,%
	reprint%
	,preprintnumbers,
	]{revtex4-1}

\usepackage{amsmath}
\usepackage{graphicx}

\begin{document}

\title{Inverse scattering problem for quantum graph vertices}

\author{Taksu Cheon$^{*}$, Pavel Exner$^{\dagger}$, Ond{\v r}ej Turek$^{*}$}


\affiliation{
{}$^{*}$Laboratory of Physics, Kochi University of Technology,
Tosa Yamada, Kochi 782-8502, Japan\\
{}$^{\dagger}$Doppler Institute for Mathematical Physics and Applied
Mathematics, Czech Technical University,
B\v rehov{\'a} 7, 11519 Prague, Czechia}

\date{\today}

\begin{abstract}
We demonstrate how the inverse scattering problem of a quantum star graph can be solved by means of diagonalization of Hermitian unitary matrix
when the vertex coupling is of the scale invariant (or F\"ul\H{o}p-Tsutsui) form. This enables the construction of quantum graphs with desired properties in a tailor-made fashion. The procedure is illustrated on the example of quantum vertices with equal transmission probabilities.
\end{abstract}

\maketitle

%

The interest in the inverse scattering problem for quantum graphs \cite{H00,BK05,GS05} is two-fold. These graphs are a prime example of solvable systems possessing nontrivial physical properties \cite{EKST08}. At the same time, it is also important for its relevance as the design principle of nanowire-based single electron devices.

In this letter, we consider the inverse scattering problem on a star graph with scale invariant vertex coupling \cite{FT00}, which is an important subset among all the couplings preserving the probability current \cite{KS99}; we note that the corresponding scattering matrix is energy independent. A star graph with a number of half-line edges connected at a single point is the elementary building block of a generic graph. In general, the inverse problem for star graphs is easy to solve \cite{Be} but the solution does not give much insight into the physical meaning of the coupling.

Here we present an alternative approach for star graphs with scale invariant coupling. We shall give the solution to the corresponding inverse scattering problem in the form of eigenvalue problem of a Hermitian unitary matrix. In particular, we consider quantum vertices with equal transmission probabilities, a subclass of the scale invariant case, and we show that the problem boils down to a search for a special class of Hermitian unitary matrices which can be regarded as a generalization of complex Hadamard matrices. Intriguing designs emerge for the realization of quantum devices with such properties.

Note that \emph{any} quantum-graph vertex coupling behaves effectively as a scale-invariant one in both the high and low energy limits \cite{CET10a}, which gives a hint that our analysis might have an extension to the general case.

Consider thus a star graph vertex of degree $n$, with $n$ half-line edges sticking out of a point-like node. The  scale invariant subfamily of the general coupling conditions is characterized by a complex matrix $T$ of size $(n-m) \times m $, where $m$ can take an integer value $m=1, 2, ..., n-1$, being given by
\begin{eqnarray}
\label{e1}
\begin{pmatrix}I^{(m)} & T \\ 0 & 0 \end{pmatrix} \Psi^\prime
= \begin{pmatrix}0 & 0 \\ -T^\dagger & I^{(n-m)} \end{pmatrix} \Psi ,
\end{eqnarray}
where $I^{(l)}$ signifies the identity matrix of size $l \times l$, and
the boundary-value vectors $\Psi$ and $\Psi^\prime$ are defined by
\begin{eqnarray}
\label{e12}
\Psi =  \begin{pmatrix} \psi_1(0) \\ \vdots \\ \psi_n(0) \end{pmatrix} ,
\quad
\Psi^\prime =  \begin{pmatrix} \psi^\prime_1(0) \\ \vdots \\ \psi^\prime_n(0) \end{pmatrix} ,
\end{eqnarray}
with $\psi_i(x_i)$ and $\psi^\prime_i(x_i)$ being the wave function and
its derivative on $i$-th edge \cite{CET10}. The coordinates $x_i$ on the $i$-th half-line are labeled outwardly from the vertex, which corresponds to $x_i=0$ for all $i$. To cast the coupling into the form (\ref{e1}) one has, in general, to renumber suitably the edges.  The quantum particle coming in from the $j$-th edge and scattered off the vertex is described by the scattering wave function on the $i$-th line, $\psi_i^{(j)}(x)$, which is of the form
\begin{eqnarray}
\label{e14}
\psi_i^{(j)}(x) =  \delta_{i,j} e^{-{\rm i} k x} + {\cal S}_{i,j} e^{{\rm i} k x} .
\end{eqnarray}
Let us express ${\cal S}$ in terms of $T$. From \eqref{e14} we have
$$
\Psi(0)=I+{\cal S}\quad\text{and}\quad\Psi'(0)=k(-I+{\cal S}) ;
$$
we substitute into (\ref{e1}), which leads to the equation
%
\begin{eqnarray}
\label{es15}
\begin{pmatrix}I^{(m)} & T \\ T^\dagger & -I^{(n-m)} \end{pmatrix} {\cal S}
= \begin{pmatrix}I^{(m)} & T \\ -T^\dagger & I^{(n-m)} \end{pmatrix} .
\end{eqnarray}
It is easy to observe from \eqref{es15} that
\begin{eqnarray}
\label{e3}
 {\cal S} = X_m^{-1} Z_m X_m ,
\end{eqnarray}
with the matrices $X_m, Z_m$ defined by
\begin{eqnarray}
\label{e3a}
 X_m = \begin{pmatrix}I^{(m)} & T \\ T^\dagger & -I^{(n-m)} \end{pmatrix},
\
 Z_m = \begin{pmatrix}I^{(m)} & 0 \\ 0 & -I^{(n-m)} \end{pmatrix}.
\ \
\end{eqnarray}
We see that (\ref{e3}) can be viewed as a diagonalization formula of Hermitian unitary matrix
 ${\cal S}$ with a diagonalizing matrix of specific block diagonal form, $X_m$, which gives a 
prescription to obtain $T$ that defines the boundary condition from the scattering matrix ${\cal S}$. In other words, solution of the our inverse scattering problem is given in terms of a diagonalization.

In practice, the procedure of obtaining $T$ by a diagonalization of ${\cal S}$ can be cumbersome for large $n$, and there is an alternative simpler way. 
A calculation shows that \eqref{e3} can be rewritten in the form
\begin{eqnarray}
\label{e404}
{\cal S} = -I^{(n)} + 2
 \begin{pmatrix}I^{(m)} \\ T^\dagger  \end{pmatrix}
\left( I^{(m)} + T T^\dagger \right)^{-1}
\begin{pmatrix}I^{(m)}  \,\ T  \end{pmatrix} .
\ \
\end{eqnarray}
 Let us divide ${\cal S}$ into four submatrices ${\cal S}_{11}$, ${\cal S}_{12}$, ${\cal S}_{21}$ and ${\cal S}_{22}$
of size $ m \times m$, $m \times (n-m)$, $(n-m) \times m$ and $(n-m) \times (n-m)$, respectively, as follows
\begin{eqnarray}
\label{e21}
{\cal S} =\begin{pmatrix}{\cal S}_{11} & {\cal S}_{12} \\ {\cal S}_{21} & {\cal S}_{22} \end{pmatrix} .
\end{eqnarray}
%
The block matrices ${\cal S}_{ij}$ express in terms of $T$ as 
$
{\cal S}_{11} = -I^{(m)} + 2 \left( I^{(m)} + T T^\dagger \right)^{-1} 
$,
$
{\cal S}_{12} =
2 \left( I^{(m)} + T T^\dagger \right)^{-1} T
$, and
$
{\cal S}_{22} = I^{(n-m)} - 2 \left( I^{(n-m)} + T^\dagger T \right)^{-1}
$. 
From here, one gets easily
\begin{eqnarray}
\label{e25}
T = \left(  I^{(m)} + {\cal S}_{11}  \right)^{-1} {\cal S}_{12}
 = {\cal S}_{21}^\dagger \left(  I^{(n-m) } - {\cal S}_{22} \right)^{-1}  .
 \quad
\end{eqnarray}
Hence, the algorithm to obtain the matrix $T$ characterizing the vertex is the following:
\begin{enumerate}
\item Take the scattering matrix ${\cal S}$ and set $m=\mathrm{rank}({\cal S}+I^{(n)})$.
\item Decompose ${\cal S}$ according to \eqref{e21}. If necessary, change the numbering of the incoming edges so that $I^{(m)}+{\cal S}_{11}$ is regular (it is always possible).
\item Calculate $T$ using \eqref{e25}.
\end{enumerate}
We remark that the matrix $T$ obtained by the algorithm above naturally depends on the numbering of the edges we choose.

In the rest of the paper we demonstrate how the matrix $T$ is used for understanding the meaning of the coupling and for the construction of the vertex with prescribed scattering properties.


Asking about the meaning of the coupling, recall a star-shaped network with a potential at the node can tend in the zero-diameter limit to the star graph with $\delta$ coupling \cite{EP09}. More complicated couplings can be obtained by using $\delta$ vertices as building blocks, applying localized magnetic fields to achieve phase change if necessary; the aim is to devise a design principle to construct an arbitrary coupling condition. Here the matrix $T$ derived above plays an important role. In the scale-invariant case (\ref{e1}) with the elements of $T=\{t_{ij}\}$, $i=1, ..., m$ and $j=m+1,.., n$, given, the scheme works as follows \cite{CT10}:

(i) Take the endpoints of the $n$ edges, numbered by $j=1, 2, ..., n$, and connect them in pairs $(i, j)$ by internal edges of length ${d}/{r_{ij}}$, except when $r_{ij}=0$ in which case the pair remains unconnected. Apply a vector potential $A_{ij}$ on the segment $(i,j)$ to produce extra phase shift $\chi_{ij}$ between the endpoints when its value is nonzero. Place the $\delta$ potential of strength $v_i$ at each endpoint $i$.

(ii) The length ratio $r_{ij}$ used above and the phase shift $\chi_{ij}$ are determined from the non-diagonal elements of the matrix $Q$ defined by
\begin{eqnarray}
\label{efcnd}
Q = \begin{pmatrix} T \\ I^{(n-m)}\end{pmatrix} \begin{pmatrix} -T^\dagger & I^{(m)}\end{pmatrix}
= \begin{pmatrix} -T T^\dagger & T \\ -T^\dagger & I^{(m)}\end{pmatrix}
 \end{eqnarray}
using the relation $r_{ij} e^{{\rm i}\chi_{ij}}=Q_{ij}$ ($i \ne j$). This means that we have $r_{ij} e^{{\rm i} \chi_{ij}} =t_{ij}$ for $i\le m, j>m$, and $r_{ij} e^{{\rm i}\chi_{ij}} = \sum_{l>m} t_{il}t_{jl}^*$ for $ i, j \le m$; for $i, j > m$  we have $r_{ij}=0$ and naturally also $\chi_{ij}=0$.

(iii) The $\delta$ coupling strength $v_{i}$ is given by the diagonal elements of the matrix $V$ defined by
\begin{eqnarray}
 \label{vicnd}
V = \frac{1}{d} (2 I^{(n)}-J^{(n)}) R,
\end{eqnarray}
where $R$ is the matrix whose elements equal the absolute values of the matrix elements of $Q$, {\it i.e.} $R=\{r_{ij}\} =\{|Q_{ij}|\}$;   the $n \times n$ matrix $J^{(n)}$ has all the elements equal to one. This means that we have
$v_{i}=\frac{1}{d}\big(1-\sum_{l \le m} r_{li} \big)$ for $i > m$,  and $v_{i}=\frac{1}{d} \big(\sum_{l > m} [r_{il}^2-r_{il}] -\sum_{l (\ne i) \le m} \!\! r_{il} \big)$ for $i \le m$.
The described way to fine tuning of the lengths and $\delta$ coupling strengths is devised to counter the generic opaqueness brought in with every addition of a vertex or a connecting edge into the graph.

The wave function $\phi(x)$$=\phi_{i,j}(x)$ on any internal edge with indices $(i, j)$ has to satisfy the relation
\begin{eqnarray}
\label{fullhlp}
\begin{pmatrix} \phi'(0) \\ e^{i \chi} \phi'(\frac{d}{r}) \end{pmatrix}
= -\frac{r}{d}
\begin{pmatrix} F(\frac{d}{r})) & -G(\frac{d}{r})  \\
 G(\frac{d}{r}) & -F(\frac{d}{r}) \end{pmatrix}
\begin{pmatrix} \phi(0) \\ e^{i \chi}  \phi(\frac{d}{r}) \end{pmatrix} ,
\ \
\end{eqnarray}
with  $F(x) = x \cot x$ and $G(x) = x {\rm \,cosec\,} x$.
Combining (\ref{fullhlp}) with the condition at the $i$-th endpoint where we have the $\delta$-potential of strength $v_i$, 
\begin{eqnarray}
\psi'_i(0)+\sum_{j\ne i} \phi'_{ij}(0) = v_i \psi_i(0)
\end{eqnarray}
we obtain the relations between the boundary values $\psi_i = \psi_i(0)$ and $\psi'_i = \psi'_i(0)$ in the form
\begin{eqnarray}
\label{full1}
d \psi'_i =
\left(
v_i d + \sum_{l\ne i}r_{il} F_{il} 
\right) 
\psi_i
\!-\! \sum_{l\ne i}  e^{{\rm i}\chi_{ij}} r_{il} G_{il} \psi_l ,
\quad
\end{eqnarray}
where the obvious notations $F_{ij}=\frac{d}{r_{il}} \cot \frac{d}{r_{il}}$ and $G_{ij}=\frac{d}{r_{il}} {\rm \,cosec\,} \frac{d}{r_{il}}$ have been adopted. Note that the equation (\ref{full1}) is exact and does not involve any approximation.
For small values of the length parameter $d$ we have $F_{ij} = 1 + O(d^2)$ and $G_{ij} = 1 + O(d^2)$; then we can show by a straightforward computation in the manner of \cite{CET10} that the shrinking limit $d \to 0$ gives the desired coupling condition for the scale-invariant vertex (\ref{e1}).

With the procedure described above,  it is possible to construct a star-graph from any given scattering matrix of the considered class. Our previous result detailed in \cite{CT10} which provides a reconstruction of the ``free-like'' scattering is one such example, and it could be achieved more easily by the current method.
Here, we will illustrate the application of the procedure on the following exemplary problem. Let us look at this question: Can one construct a quantum vertex for which the particle incoming from {\it any} line is transmitted to all other lines with equal probability? At first, we should ask about the existence condition for the scattering matrix of the form
\begin{eqnarray}
\label{ee01}
{\cal S} =
\frac{1}{\sqrt{d^2+n-1}}
\begin{pmatrix}
d & \!\!\!\! e^{ {\rm i} \phi_{12}} &\!\!\!\!  &\!\!\!\!\!\!\!\! \cdots &\!\!\!\! e^{ {\rm i} \phi_{1n}} \\
e^{ {\rm i} \phi_{21}} &\!\!\!\! d &\!\!\!\!   &\!\!\!\!\!\!\!\! \cdots &\!\!\!\! e^{ {\rm i} \phi_{1n}} \\
\vdots &\!\!\!\! &\!\!\!\!\ddots&\!\!\!\!\!\!\!\!  &\!\!\!\! \vdots \\
e^{ {\rm i} \phi_{n-11}}&\!\!\!\! \cdots &\!\!\!\! &\!\!\!\!\!\!\!\!-d&\!\!\!\! e^{ {\rm i} \phi_{n-1n}} \\
e^{ {\rm i} \phi_{n1}} &\!\!\!\!  \cdots &\!\!\!\! &\!\!\!\!\!\!\!\! e^{ {\rm i} \phi_{nn-1}} &\!\!\!\! -d\\
\end{pmatrix} ,
\quad
\end{eqnarray}
with a non-negative real parameter $d$. 

These matrices have been recently examined in \cite{TCxx}. It has been proved that the parameter $d$ is always bounded from above by $\frac{n}{2}-1$ (except for $n\leq2$), and moreover, that for most values of $d\in[0,\frac{n}{2}-1]$, the existence of the corresponding matrix \eqref{ee01} is impossible if the order $n$ is odd. By contrast, if $n$ is even, then one can construct an ${\cal S}$ for infinitely many values of $d$, in particular for any $d$ from the interval $[\frac{n}{2}-3,\frac{n}{2}-1]$, or, under certain extra condition, for all $d\in[0,1]$; for further details and explicit matrix constructions we refer to \cite{TCxx}.
%
\begin{figure}
\center{
\includegraphics[width=4.0cm]{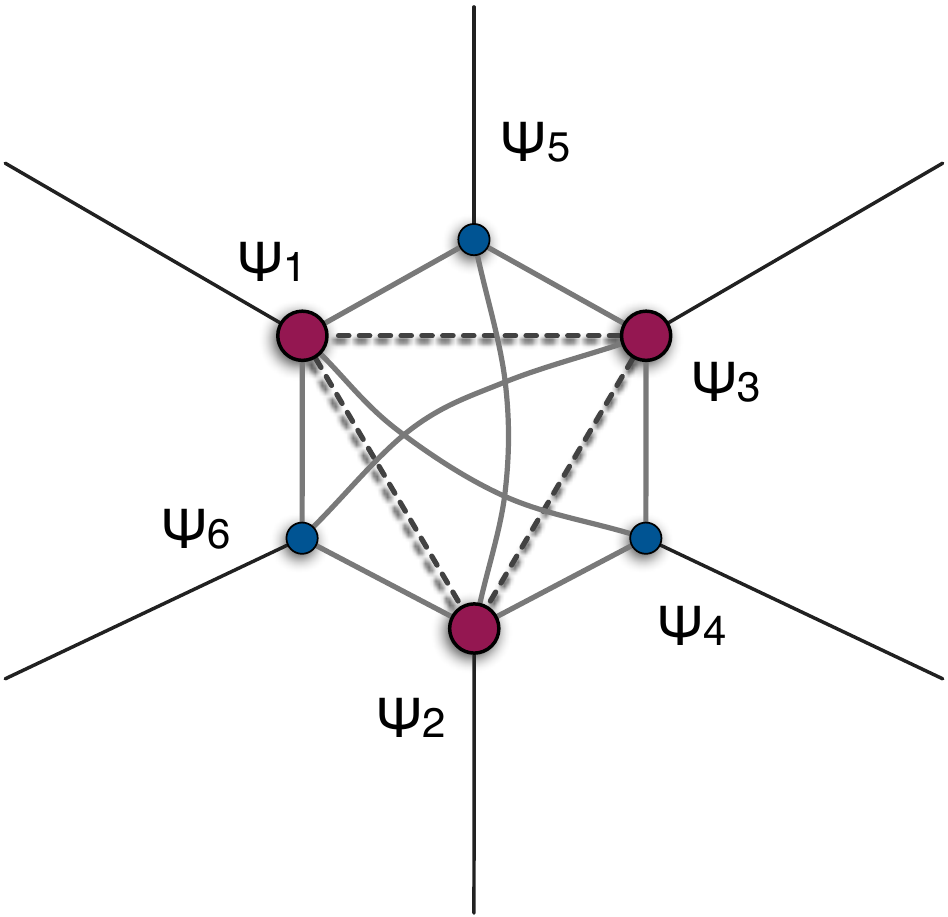} \
\includegraphics[width=4.0cm]{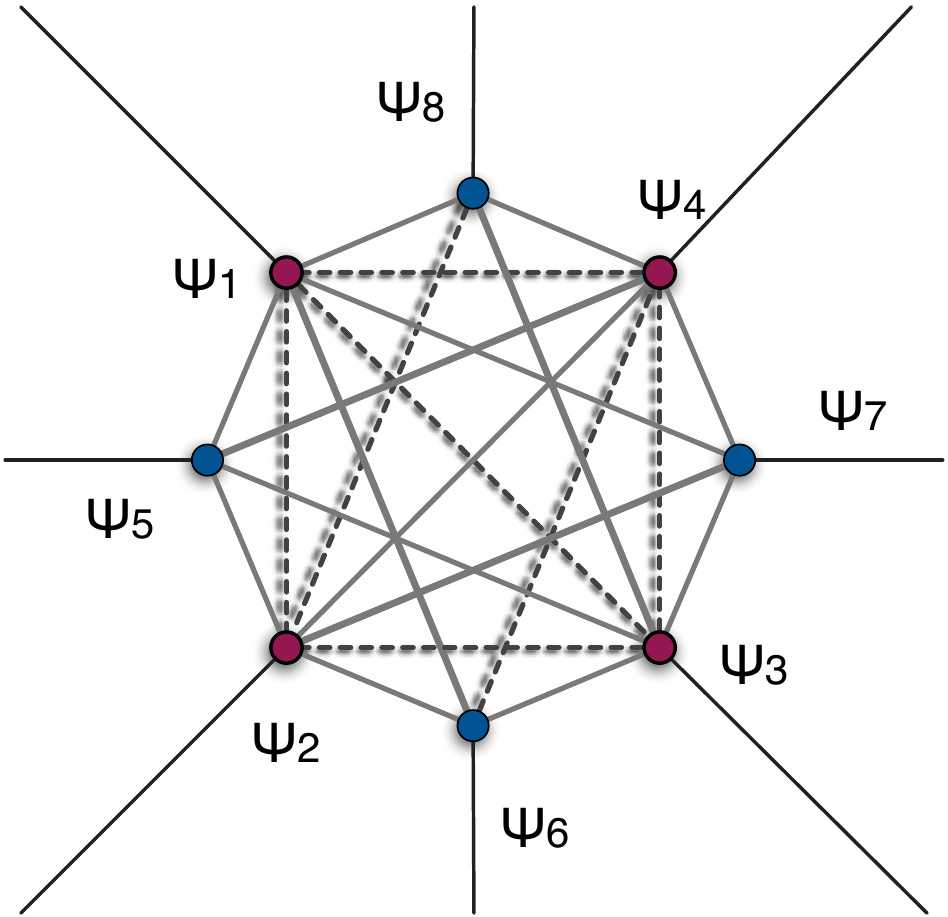}
}
\label{fig1}
\caption
{
Finite approximation to the reflectionless
scale invariant vertices corresponding to (\ref{scc06}) (left)
and (\ref{es08})  (right) constructed according to (\ref{efcnd})--(\ref{vicnd}). Dotted line indicates the presence of a non-zero phase shift $\chi_{ij}$.
}
\end{figure}

With a matrix of the type \eqref{ee01} in hand, we can proceed to the construction of the finite approximation. We will demonstrate it on two examples. We first look at the case of a reflectionless scattering with uniform transmission to all the other edges. In other words, we ask whether there is a graph with the scattering matrix \eqref{ee01} with $d=0$. 
Vertices yielding such scattering matrices are known to be useful in investigation of semiclassical properties of quantum-graph spectra \cite{HS07}.
%
Let us adopt a result from \cite{TCxx} which says that such an ${\cal S}$ can exist only for even $n$. We limit ourselves for simplicity to real ${\cal S}$; we remark that a real matrix of the type \eqref{ee01} with zero diagonal is called {\it symmetric conference matrix}, and is known to exist for $n=2, 6, 10, 14, 18, 26, 30, 38, ...\:$.

Here we inspect the example of $n=6$ when ${\cal S}$ is given by
\begin{eqnarray}
\label{scc06}
{\cal S} =  \frac{1}{\sqrt{5}}
\begin{pmatrix}
I^{(3)}-J^{(3)} & -2I^{(3)}+J^{(3)} \\
-2I^{(3)}+J^{(3)} & -I^{(3)}+J^{(3)}
\end{pmatrix} .
\end{eqnarray}
Applying \eqref{e25}, the corresponding $T$ is easily calculated,
\begin{eqnarray}
\label{scc06t}
{T} = -\gamma I^{(3)}+(1+\gamma)J^{(3)}
\end{eqnarray}
where $\gamma = (\sqrt{5}-1)/2$ is the golden mean.
Our finite approximation is specified by following parameters;
%
$
r_{12}=r_{23}=r_{13}=4+3\gamma,
$
$
r_{14}=r_{25}=r_{36}=1,
$
$r_{15}=r_{16}=r_{26}=r_{24}=r_{31}=r_{32}=1+\gamma
$,
$
r_{45}=r_{46}=r_{56}=0
$,
$e^{ {\rm i} \chi_{12}}=e^{ {\rm i} \chi_{23}}=e^{ {\rm i} \chi_{13}}=-1
$, 
$e^{ {\rm i} \chi_{ij}}=1$
for all others, and
$v_1 = v_2 = v_3 = -6\frac{\gamma+1}{d}
$,
$
v_4 = v_5 = v_6 = -2\frac{\gamma+1}{d} 
$.
%
The finite graph approximation is schematically illustrated in the left side of Figure 1.

Second example is the equal-scattering graph,
in which the scattering is uniform to all the edges
including the one of the incoming particle.
Such a matrix, called {\it symmetric Hadamard matrix}, is known to exist for $n=2, 4, 8, 12, 16, 20, 24, ...\:$.
An example of such ${\cal S}$ for $n=8$ is given by
\begin{eqnarray}
\label{es08}
{\cal S} = \frac{1}{\sqrt{8}}
\begin{pmatrix}
2I^{(4)}-J^{(4)} & -2I^{(4)}+J^{(4)} \\
-2I^{(4)}+J^{(4)} & -2I^{(4)}+J^{(4)}
\end{pmatrix} .
%
\end{eqnarray}
The matrix $T$ specifying
the vertex coupling is found to be equal to
\begin{eqnarray}
\label{es08t}
T = 
\frac{\sigma-1}{\sigma+1} I^{(4)}+ \frac{1}{\sigma+1} J^{(4)}
\end{eqnarray}
where $\sigma=\sqrt{2}-1$ is the silver mean. Our finite approximation is specified by following parameter values;
%
$
r_{12}=r_{13}=r_{14}=r_{23}=r_{24}=r_{34} =1+\sigma
$,
$
r_{15}=r_{26}=r_{37}=r_{48}=\frac{\sigma}{1+\sigma}
$,
$r_{16}=r_{17}=r_{18}=r_{27}=r_{28}=r_{25}
=r_{38}=r_{35}=r_{36}=r_{45}=r_{46}=r_{47}=\frac{1}{1+\sigma}
$,
$
r_{56}=r_{57}=r_{58}=r_{67}=r_{68}=r_{78}  =0
$,
$
e^{ {\rm i} \chi_{12}}=e^{ {\rm i} \chi_{13}}=e^{ {\rm i} \chi_{14}}
=e^{ {\rm i} \chi_{23}}=e^{ {\rm i} \chi_{35}}
$
$
=e^{ {\rm i} \phi_{28}}=e^{ {\rm i} \chi_{46}}
=-1
$,
$e^{ {\rm i} \chi_{ij}}=1 {\rm \ for \ all \ others}
$, and
$
v_1 = v_2 = v_3 = v_4 =  -\frac{5\sigma+3}{d}
$,
$v_5 = v_6 = v_7 = v_8 =  -\frac{\sigma+1}{d}
$.
%
The finite graph approximation is schematically illustrated in the right side of Figure 1.

Finally, let us discuss the convergence of the the described finite-size graph approximation. In Figure 2, we display scattering matrix elements of the finite graph  constructed to approximate the equal-scattering reflectionless matrix (\ref{scc06}). They are calculated directly from (\ref{full1}). The scale of the wave length $k$ is given by $1/d$. The approximation can be seen to be quite good below $kd < 0.2$. Numerical analysis of other examples of different graphs gives essentially the same conclusion, namely that
the described construction does represent a physical realization of scale-invariant vertex couplings.

\begin{figure}
\center{
\includegraphics[width=6.0cm]{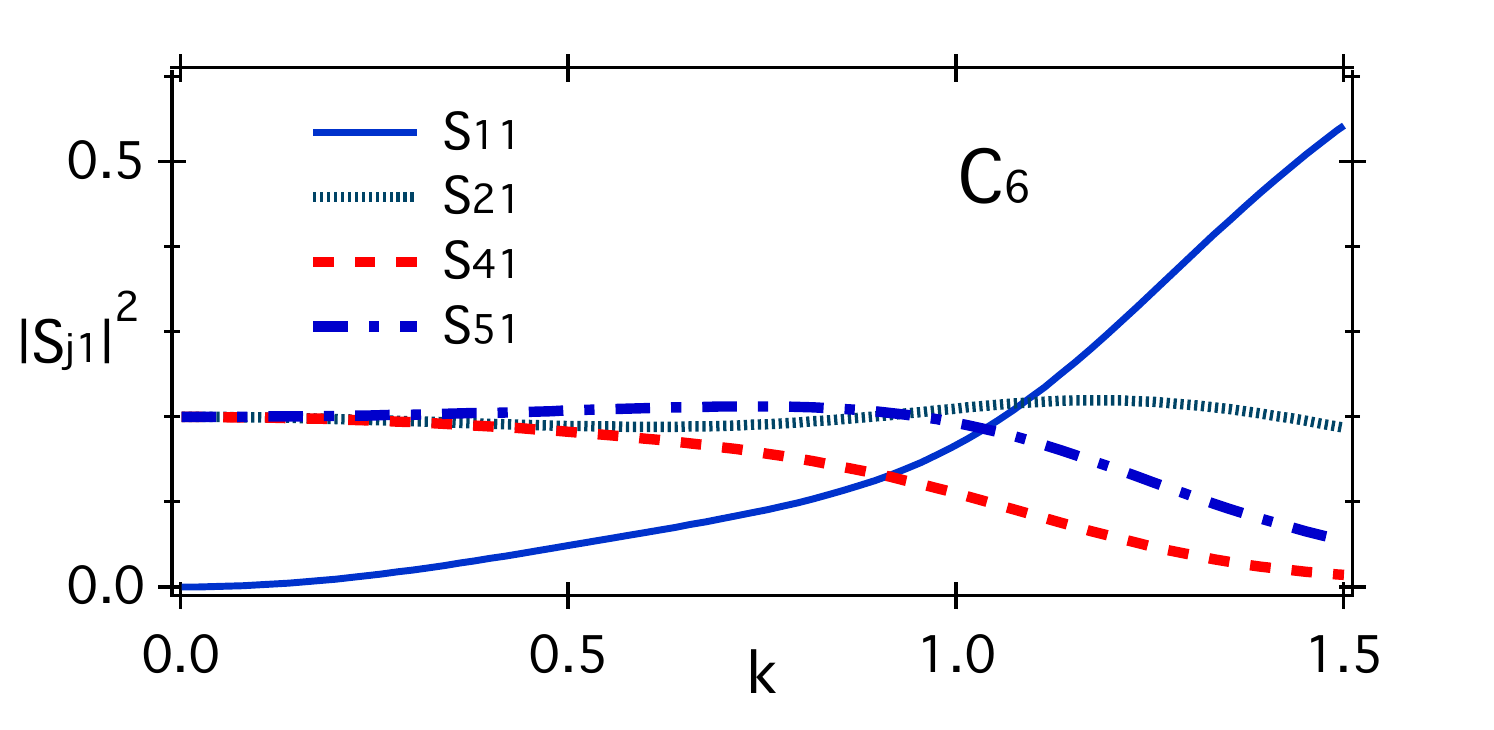}}
\center{
\includegraphics[width=6.0cm]{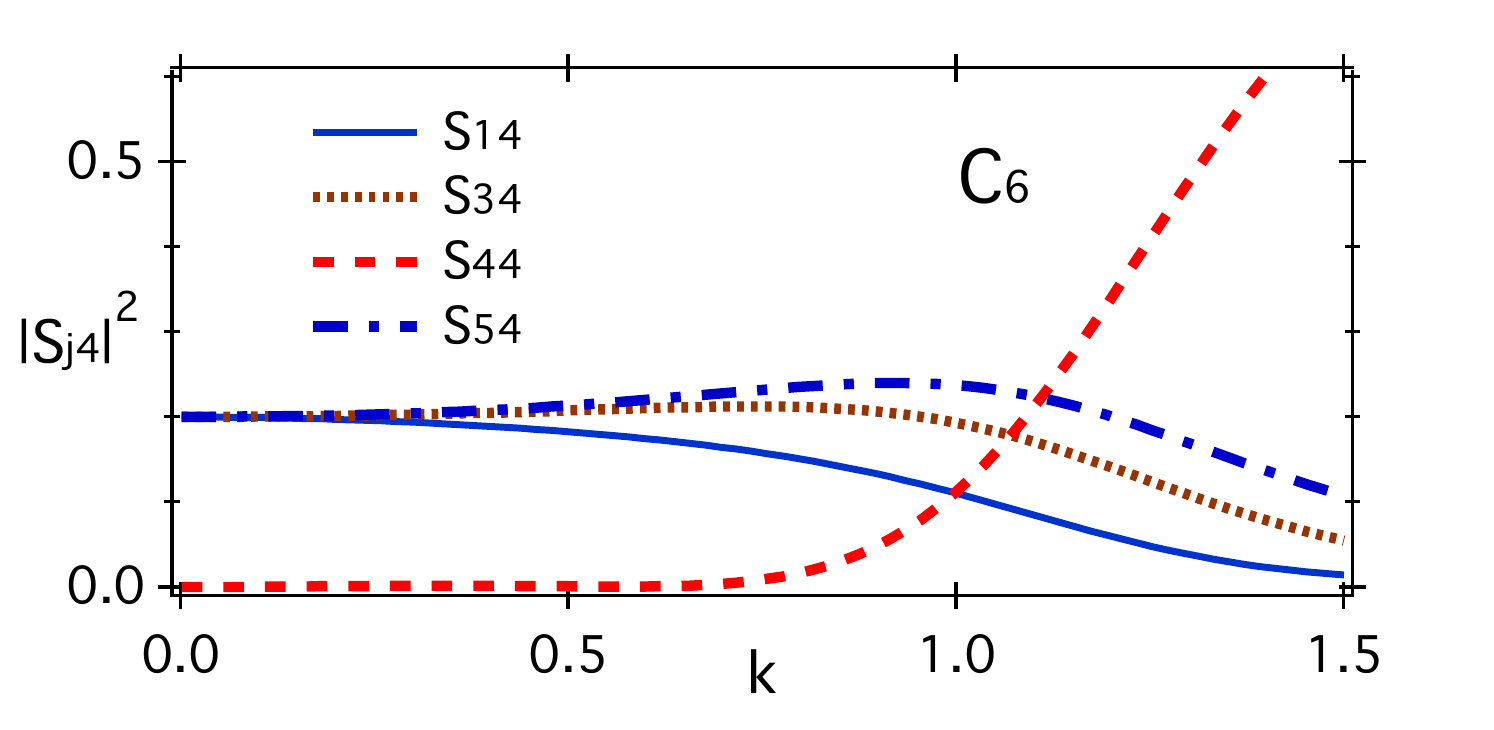}}
\label{fig2}
\caption
{
Scattering probabilities as functions of incoming momentum $k$ (in the unit of $1/d$) of finite quantum graph approximating
the equal-transmitting reflectionless vertex represented in the left side of Figure 1.
}
\end{figure}

Thus the problem of finding the desired property of  scale invariance is turned into a mathematical question about a Hermitian unitary matrix, and the search for systems with ${\cal S}$ having interesting specifications other than those examined here should follow. 
Also, a study of the bound state spectra is one thing we have completely neglected in this letter; 
applications to non-quantum waves, including particularly electromagnetic and  water waves should be another interesting subject.

In our finite approximation of star graph with no internal edges, we have actually studied the low-energy properties of graphs with internal edges all of which are connected to the external ones, which we might term {\it depth-one} graphs. The examination of {\it depth-two} graphs
and beyond seems to be a natural future direction.
Our result showing the full solution to the inverse scattering problem is, in a sense, a partial fulfillment of the hope that quantum graph somehow could be a solvable model and useful design tool at the same time.

\medskip
We thank Prof. U. Smilansky, Prof. L. Feher, Prof. I. Tsutsui, Prof. T. Kawai 
and Prof. T. Shigehara
for stimulating discussions. This research was supported  by the Japanese Ministry of Education, Culture, Sports, Science and Technology under the Grant number 21540402, and also by the Czech Ministry of Education, Youth and Sports within the projects LC06002 and P203-11-0701.

\end{document}